# Defending the future: An MSc module in End User Computing Risk Management

Simon Thorne
SThorne@uwic.ac.uk
University of Wales Institute Cardiff

**Abstract**

This paper describes the rationale, curriculum and subject matter of a new MSc module being taught on an MSc Finance and Information Management course at the University of Wales Institute in Cardiff. Academic research on spreadsheet risks now has some penetration in academic literature and there is a growing body of knowledge on the subjects of spreadsheet error, human factors, spreadsheet engineering, "best practice", spreadsheet risk management and various techniques used to mitigate spreadsheet errors. This new MSc module in End User Computing Risk Management is an attempt to pull all of this research and practitioner experience together to arm the next generation of finance spreadsheet champions with the relevant knowledge, techniques and critical perspective on an emerging discipline.

**1.0 Introduction**

When designing any course, module or even lecture material, similar questions arise. The first being: What should be included in the syllabus? – given the wealth of research on spreadsheets, which areas should be emphasised? One then has to consider specifically what topics and subjects should be covered. Finally one has to decide how to prepare the material in terms of lecture and seminar mix for the students.

My philosophy when designing this module is to assume nothing about the students coming into the module. In a similar way that an organisation may become suddenly aware of its spreadsheet dependency, the material in this module is designed to make the student aware of the pervasive nature of spreadsheet applications, the problems they carry and the current best thinking on how to manage the risk they pose.

Of course this raises some interesting questions, one might be able to define risk quantitatively or qualitatively by some process, but only relative comparisons can be drawn from such analysis – i.e. without context the "risk" of a spreadsheet can only be compared against other spreadsheets rated in the same manner . Also there are several contentious issues in spreadsheet research such as "best practice". As Colver (2004) notes, best practice in one discipline is often problematic in another. Hence there are no silver bullets for spreadsheet design but there are nuggets of wisdom in various sub disciplines that could make the job of managing and controlling spreadsheet use in organisations easier and safer.

Of course, any attempt to control spreadsheet use in an organisation must come from a risk management perspective, i.e. spreadsheets are used in many different departments in organisations as observed by SERP (2006) but not all of these spreadsheets will need managing. Indeed, studies show that organisations posses so many spreadsheet applications, it would be infeasible to audit every single spreadsheet (Butler 2000 and Pryor 2004).

So one must identify the critical and risky spreadsheets and do something about them which of course implies that you have a metric or means of measuring spreadsheet risk. Work such as Madahar *et al* (2007) provides some insight and method for this approach, indeed this is the most up to date research in spreadsheet risk management of this type. Other approaches exist such as Baxter (2004) who proposes a decentralised technological framework for controlling the use and version control of spreadsheet applications. However, this approach is not led by a risk assessment metric, as Madahar *et al.* (2007) propose, rather this technique is based on the alteration of critical spreadsheets through an agent based surveillance system.

**2.0 Background**

A joint effort between the Finance & Accounting and Information Systems & Computing department at the University of Wales Institute Cardiff (UWIC) saw the validation of a new MSc in Finance and Information Management in late 2009. This programme is aimed at the next generation of financial decision makers who need an appreciation of both finance and information management.

**2.1 MSc Finance and Information Management**

To set the context for the rest of this paper a brief synopsis of the course and its structure and content of the modules will be provided.

The course is designed for financial decision makers who need an appreciation of finance and information management, therefore the makeup of the MSc is a mix of finance and Information Systems (IS) modules.

The course is made up of 7 taught modules: Quantitative finance; Capital Markets and derivatives; Corporate finance and risk; End User Computing (EUC) risk management; Analysis and design; Information management software tools and Research methods. The students then produce a dissertation as the final assessed element of the course.

The IS modules taught as part of this programme complement each other in several ways. The 'analysis and design' module equips the student with the skills necessary to conceptually model business problems using a variety of structured techniques. The rationale for this module is that since financial decision makers are likely to be designing and building their own information systems, this module will provide them will the skills necessary to define and plan the models conceptually before they are implemented.

Knowledge of conceptual modelling processes will assist students in planning and modelling spreadsheets effectively. Although the techniques are not tailored to spreadsheets specifically, the skill of conceptual modelling is beneficial when writing spreadsheet applications (Rajalingham *et al*, 2000). Also, work in human factors and cognitive science suggests that improving a spreadsheet developer's skill in basic logic has a positive impact on the quality of spreadsheet model they produce (Kruck *et al*. 2003).

The 'Information management software tools' module instructs the students in managing information via a variety of software tools. For instance, one part of this module deals with implementing queuing theory models in software, i.e. determining
how resources should be allocated (such as staff costs) given a demand on a service. This module focuses on the actual development of software for such models rather than the operations research theory.

Both the 'analysis and design' and 'information management software tools' complement the end user computing risk management module in several ways. The information management software tools module is the practice of creating software for managing real world information. The students will be able to take what they have learnt in analysis and design to plan the work and the knowledge gained from end user computing risk management to understand and manage the inherent risks such spreadsheet activities carry.

**2.2 End User Computing Risk Management**

Financial decision makers use information systems to manage, analyse and support decision making in a financial setting. We know from surveys and case studies that the finance industry is reliant on spreadsheet applications (Croll, 2005).

We also know from spreadsheet research that the use of spreadsheet applications is risky (Panko, 1999) and that spreadsheet developers are prone to a spectrum of difficulties when developing spreadsheet models which may result in errors which may ultimately result in significant financial loss (French, 2003).

Therefore, the next logical step to take in defending the future of organisations from spreadsheet risks is to educate the next generation of financial spreadsheet developers and managers with the relevant critical appreciation of the subject, the practical techniques for ensuring spreadsheet quality (or at least reducing defects) and the skills to enable them to effectively manage and control spreadsheet development in their organisation. Research consistently shows that spreadsheet developers have little or no training (Taylor *et al.* 1998, Gosling 2003, Fernandez 2003, SERP 2006) let alone 'education' on managing spreadsheet risks. This module is an attempt to address this skills gap from a risk management perspective.

Whilst spreadsheet research is still an emerging research focus, the work of spreadsheet researchers, practitioners and organisations such as EuSpRIG has now yielded a wealth of information, knowledge and techniques concerning spreadsheet risks.

**2.2.1 Learning outcomes of the module**

The specific learning outcomes of the module are as follows:

- Critically evaluate the role of EUC in modern organisations
- Demonstrate understanding of the risks associated with EUC, specifically spreadsheet error.
- Understand the role Human Factors plays in spreadsheet development.
- Evaluate and discuss methods applied to spreadsheets to reduce errors presented in academic research.
- Formulate suitable management strategies for management of EUC in organisations
- Understand the role of legal, management and technological policies for the control of EUC in organisations.

Having met these learning outcomes, the student should be sufficiently prepared to assist organisations with their spreadsheet development and control.

**2.3 Curriculum**

The curriculum of subjects to be covered during the year is as follows: Information systems in organisations, end user computing; spreadsheet errors and risks, risk management techniques, spreadsheet risk management, spreadsheet error management and control techniques.

The lecture and seminar material is split into blocks that cover certain broad topics that in turn address particular subjects within that topic. See table 1 below for a summarised break down of each subject area.

| Block | Lectures | Workshop tasks |
|---|---|---|
| **Introduction** | IS in organisations.<br>(1 Week) | Background reading. |
| | End User Computing.<br>(2 Weeks) | Discussions of EUC management literature. |
| **Spreadsheet risks** | Spreadsheets.<br>(1 week) | Spreadsheet basics. |
| | Spreadsheet errors.<br>(3 weeks) | Repetition of various spreadsheet creation experiments such as the wall and ball task to highlight the problem with spreadsheets. |
| | Taxonomies of error and human factors.<br>(3 Weeks) | Manual auditing methods, code inspection, Base Error rate, overconfidence |
| | Error reduction techniques and spreadsheet engineering.<br>(3 Weeks) | Various check and control techniques, auditing experiments, "best practice" and peer audit. |
| **Risk Management** | Introductory Risk Management.<br>(2 Weeks) | Risk management literature. |
| | The law and spreadsheets.<br>(2 Weeks) | Sarbanes Oxley, Basil and others. |
| | Spreadsheet risk management.<br>(4 Weeks) | Arriving at categories of use and relative risk. |
| | Implementing Spreadsheet risk management strategies.<br>(2 Weeks) | Case studies such as Gosling 2003 & Fernandez 2003, Chambers 2008. |

**Table 1 EUC Risk Management syllabus**

This brief outline covers all of the lecture material, the tutorial material (practical tasks associated with the lectures) is more difficult to define. However, pulling on previous work such as Chadwick (2003) who defined the T.E.A.M approach – Tools Audit Education Management, Vandeput (2009) on spreadsheet education, Read and Batson (1999) on spreadsheet modelling best practice, one can formulate an essential set of anti spreadsheet error skills that the financial modeller of the future should be armed with.

The first block of lectures (see table 1) is concerned with setting the scene for End User Computing (EUC) in the modern organisation. Most of the students will have a limited understanding of how information systems support the role of modern organisations, so this is the subject of the first lecture. This lecture will conclude with a look at how the term End User Computing was coined and an examination of what the term means today. The next lecture will then examine what constitutes end user computing in terms of software artefacts, development tools and the relative benefits and limitations of such activities. This lecture will conclude with an examination of the most prevalent EUC artefact, the spreadsheet.

The next block of lectures will examine: Spreadsheet errors, spreadsheet error types, human factors and spreadsheet error reduction techniques. The emphasis in this block will be to underline the prevalence and nature of spreadsheet errors and how some of the error reduction techniques can be used to limit the number of defects in spreadsheets.

The next block of lectures will introduce risk management as a concept through some introductory literature. We will then move on to consider the spreadsheet risk management research that currently exists and examine how these practices might be incorporated into organisations.

**2.4 Assessment**

The assessment in this module takes place through a significant piece of coursework and an end of year exam. The coursework will require the student to recommend a course of action regarding spreadsheet management for an organisation based upon a case study provided to them. The case study will provide all the necessary information on the organisations background and current spreadsheet activities. The actual case study will be generated from a number of case studies presented at EuSpRIG and other research resources. For example, elements of Chambers and Hamil (2008) Gosling (2003) and Fernandez (2003) will be incorporated into a new case which the students will critically appraise and recommend suitable practices and policies for controlling the organisations end user computing activities.

**2.5 We're all human**

Whilst the main focus of the module is risk and its management, part of managing risk is first hand experience of the errors and practices that make spreadsheet use risky. This module is designed to take the students on a journey as they progress through the module. Wherever possible this journey will be a 'bottom up' approach, for example allowing the students to create the mistakes when writing spreadsheets, see the impact of those mistakes and to attempt to reduce the errors through auditing and testing. This approach is preferable since most students see spreadsheets as something they mastered when studying at secondary school. Some anecdotal evidence to this effect was observed when conducting fieldwork for my doctoral thesis with Masters students. The students were acting as participants in some spreadsheet error research, after completing the exercise and being asked for feedback, one student said:

*"The exercise was pointless, my father is an accountant and I have worked for the family business for 10 years. There is nothing I don't know about spreadsheets and I know my spreadsheet models are correct"*

Participants were anonymous in the task, so it was impossible to trace the exact performance of this student. However, none of the participants achieved 100% which means that the student in question must have made some mistakes.

Of course this is a classic case of overconfidence (Panko, 2003) and one that is probably typical amongst the students. Whilst many of them won't have ten years experience working for an accounting firm, they will probably see writing spreadsheet models as a basic skill and one that can be implemented without any training. To mitigate this inherent overconfidence, a bottom up strategy will be employed. Get the students to: participate in creating errors, appreciate how easily mistakes are made and then experience the difficulty of spotting and correcting them. It is hoped that this technique will allow the students to develop a critical understanding from the bottom up – not only will they understand the management issues but also the issues that are critical when creating spreadsheet models.

The End User Computing Risk Management module runs as a "year long" module (meaning that students will start the module in semester 1 and finish in semester 3). This allows the module to be delivered over a longer time period which is beneficial since it is likely that most of the students will have no experience in spreadsheet risk management. It also allows the lecturer to pace the delivery of information in a gentler fashion – one can really set the scene for the module without the worry of over-burdening the students.

**2.6 The future of spreadsheet education**

Designing and running a module in spreadsheet risk management is perhaps the first step towards dedicated postgraduate and undergraduate courses in spreadsheet risk and engineering.

At present, spreadsheet engineering is an emerging discipline that has some clear definitions and practices but one that is yet to be fully realised. The work of several researchers (Burnett et al 2004, Grossman 2002, Grossman and Ozluk 2004, Panko 2006, Rust et al 2006, Pryor 2004, Yirsaw et al 2003, Rajalingham 2000) and others have all contributed to the growing spreadsheet engineering discipline but at present this work collectively lacks cohesion, perhaps risk management could be the solution to this problem – one can use a risk management strategy to bring all of this work together, i.e. in the case of a spreadsheet being X risk, apply Y technique. Much in the way that Madahar *et al.* (2007) propose. This is of course a top down approach to the problem, and one must consider the alternative which is to give spreadsheet modellers themselves the ability to engineer spreadsheets to get "good enough spreadsheets", which is what eventually happened in mainstream software engineering post software crisis.

It can only be a matter of time before spreadsheet engineering is recognised as a discipline within the academic and practitioner world. As was the case of the software crisis and the eventual emergence of software engineering, the world will eventually realise that spreadsheets can be engineered to be safer and that the careful application of such techniques does not stifle the creativity and flexibility of spreadsheet modellers.

Perhaps a new role in organisations is emerging, the spreadsheet engineer, the spreadsheet champion of the office who assists colleagues to write spreadsheets and ensure their quality. But one who has a wider appreciation of the problems with spreadsheet quality and knows how to manage the risk of such activities. A combined bottom up (spreadsheet administrator) and top down (spreadsheet risk management) approach might be enough to control spreadsheet use in organisations.

It is also worth noting the changing technological environment. The computer industry is changing, Apple is now the most valuable tech company in the world, this current position seemed impossible 10 years ago. Computers are getting smaller and smaller, how long will it be before major decisions will be made on train journey using a spreadsheet on an iPhone or an iPad? Will there be a new class of end users writing end user applications on these devices and what benefits and risks do these bring?

Of course it is impossible to predict how society will use these new powerful compact computers precisely but one can be mindful of a changing environment and engage with it wherever possible.

**3.0 Conclusions**

This paper has presented a brief overview of a new MSc module in spreadsheet risk management, the rationale for such a programme, the content of the module, the assessment and some of the inspiration and thinking behind such a venture.

This module is a result of researching spreadsheet risks and engaging with the community over the last 8 years. It represents the work of many academics and practitioners who have participated in EuSpRIG and other spreadsheet focussed groups such as SPRIG and the spreadsheet mini track at HICSS.

The most difficult aspect of designing such a programme is capturing and communicating the imagination and hard work involved from the spreadsheet community.

In conclusion, this module is only the beginning of an emerging spreadsheet discipline which will in time have entire postgraduate and undergraduate courses devoted to it.